\documentclass{llncs}
\usepackage{graphicx}
\usepackage{multirow}
\begin{document}
\title{Design of SEC-DED and SEC-DED-DAEC Codes of different lengths}
\author{Sayan Tripathi\and
Jhilam Jana \and
Jaydeb Bhaumik}
\institute{Department of Electronics and Telecommunication Engineering \\
	Jadavpur University, Kolkata\\
\email{tripathysayan@gmail.com, jhilamjana2014@gmail.com, bhaumik.jaydeb@gmail.com}}
\maketitle              
\begin{abstract}
Reliability is an important requirement for both communication and storage systems. Due to continuous scale down of technology multiple adjacent bits error probability increases. The data may be corrupted due soft errors. Error correction codes are used to detect and correct the errors. In this paper, design of single error correction-double error detection (SEC-DED) and single error correction-double error detection-double adjacent error correction (SEC-DED-DAEC) codes of different data lengths have been proposed. Proposed SEC-DED and SEC-DED-DAEC codes require lower delay and power compared to existing coding schemes. Area complexity in terms of logic gates of proposed and existing codes have been presented. ASIC-based synthesis results show a notable reduction compared to existing SEC-DED codes. All the codec architectures are synthesized on ASIC platform. Performances of different SEC-DED-DAEC codes are tabulated in terms of area, power and delay.\\
\textbf{keywords:} {Error correction codes (ECC), Soft errors, SEC-DED, SEC-DED-DAEC, ASIC}
\end{abstract}
\section{Introduction}
Nowadays, consumer demand for more functionalities, low power consumption and compact systems. Memory is an important part of many electronics gadgets. The major concern for memories is soft errors which are caused by radiation \cite{Dixit2011}, \cite{Ibe2010}. These soft errors corrupt the digital data and multiple bit upsets (MBUs) have occurred. So to impart more reliability to these systems, errors must be detected and corrected. Several error detecting and correcting codes are already available. Many adjacent error correcting codes \cite{Hamming1950}, \cite{Hsiao1970}, \cite{Dutta2007} and CA-based error detecting and correcting codes \cite{Bhaumik2010}, \cite{Samanta2015}, \cite{Samanta2018} have already been introduced to detect and correct adjacent errors in communication and storage systems. Alternatively, Bose-Chaudhuri-Hocquenghem (BCH) code \cite{Zhang2018} and Reed Solomon (RS) code \cite{Rev2015}, \cite{Samanta2017} can protect MBUs.  \\
Cha and Yoon proposed a technique to design an ECC processing circuits for SEC-DED code in memories. The area complexity of the ECC processing circuits have been minimized in \cite{Cha2012}. Adalid et al. presented a SEC-DED code for short data words which can detect the double bit errors and correct single error \cite{Adali2016}, \cite{Adalid2016}. \\
Alabady et al. proposed a coding technique to detect and correct single and multiple bit errors in \cite{Alabady20182}, \cite{Alabady20183}, \cite{Alabady20181}. The algorithms, flowchart, error patterns and its syndrome values are presented in \cite{Alabady20182}, \cite{Alabady20183}, \cite{Alabady20181}. Alabady et al. codes are unable to satisfy single error correction and double error detection functionality in some cases. Beside this limitation, there are some mistakes in flowchart, tables and figure which are rectified in \cite{Tripathi2018}. Ming et al. proposed a SEC-DED-DAEC code to diminish noise source in memories \cite{Ming2011}. These existing codes require more area, power and delay.\\
To mitigate these problems, this paper aims to develop new channel coding techniques. This work proposes a modified SEC-DED-DAEC code for memories. Also, this paper identifies the mistakes in proposed $H$-matrix construction procedures, the formation of equation 6 and one table of ref. \cite{Tripathi2019}. These typos do not affect the main contributions and results of the paper in \cite{Tripathi2019}. Here we have corrected these mistakes in \cite{Tripathi2019}. The main contributions are as follows:\\
i) New method to construct the parity check matrices $(H)$ for SEC-DED-DAEC code has been proposed. ii) SEC-DED-DAEC codes with different message length have been designed and implemented in ASIC platform and iii) proposed codes are fast and power efficient compared to existing designs. The rest of this paper is organized as follows. Section II provides design of proposed SEC-DED-DAEC codes. Section III presents estimation of logic gates for different designs. Section IV contains synthesis results and Section V presents the conclusion.
\section{Design of Proposed SEC-DED-DAEC Codes}
Proposed $(n-k)$ error correction code is a linear block code with parity check matrix $(H)$ which consists of $(n-k)$ number of rows and $n$ number of columns.  There are some mistakes in the construction procedure of (14, 8) proposed $(H)$ matrix in ref. \cite{Tripathi2019}. In this section the corrected construction procedure of proposed $(H)$ matrices for both SEC-DED and SEC-DED-DAEC codes with different message lengths has been described.
\subsection{$H$-matrix construction procedures}
The procedure to generate the (14, 8) proposed $H$-matrix for both SEC-DED and SEC-DED-DAEC codes is as follows:\\
\textbf{Step 1}: The $H$-matrix consists of $(n-k)$ number of rows and $(n)$ numbers of columns with $k$ numbers of data columns and $(n-k)$ numbers of parity columns having identity property.\\
\textbf{Step 2}: Last data column $(d_8)$ has been selected to satisfy the weight 3 as well as modulo-2 sum of $d_8$ and parity column $(p_1)$ will generate `1' in positions 1, 3, 4 and 6.\\
\textbf{Step 3}: Data column $(d_7)$ is selected to satisfy the weight 3 as well as modulo-2 sum of $d_7$ and data column $(d_8)$ will generate `1' in positions 1, 2, 4 and 6.\\
\textbf{Step 4}: Process is continued up to the first data column $(d_1)$ using following $Q$-matrix with target to reduce delay and power consumption without violating $H$-matrix construction rules.\\
\begin{figure}[]
\centering
\[
Q=
  \left[ {\begin{array}{cccccccccccccc}
1	1 1 3 1 2 1 1\\
2	3	4	4	2	3	2	3\\
4	5	5	5	3	4	4	4\\
5	6	6	6	4	6	6	6\\

  \end{array} } \right]
\]
\caption{$Q$-matrix of (14, 8) proposed SEC-DED and SEC-DED-DAEC codes}
\label{fig1}
\end{figure}
The $H$-matrix of (14, 8) SEC-DED and SEC-DED-DAEC codes are obtained by employing the proposed $H$-matrix construction methodology and it is shown in Fig. \ref{fig1}. This $H$-matrix consists of 8 data columns and 6 parity columns as shown in Fig. \ref{fig2}.
\begin{figure}[]
\centering
\[
H=
  \left[ {\begin{array}{cccccccccccccc}
1	0	1	0	0	1	1	0	1	0	0	0	0	0\\
0	1	1	1	1	0	1	0	0	1	0	0	0	0\\
1	1	0	0	1	0	1	1	0	0	1	0	0	0\\
1	0	0	1	0	1	0	1	0	0	0	1	0	0\\
0	1	0	1	0	0	0	0	0	0	0	0	1	0\\
0	0	1	0	1	1	0	1	0	0	0	0	0	1\\

  \end{array} } \right]
\]
\caption{$H$-matrix of (14, 8) proposed SEC-DED and SEC-DED-DAEC codes}
\label{fig2}
\end{figure}
Similarly the other $H$-matrices are constructed employing proposed construction procedures.
\begin{figure}[]
\centering
\[
H=
  \left[ {\begin{array}{cccccccc}
1	1	0	1	0	0	0	0\\
0	1	0	0	1	0	0	0\\
0	1	1	0	0	1	0	0\\
1	0	1	0	0	0	1	0\\
1	0	1	0	0	0	0	1\\

  \end{array} } \right]
\]
\caption{$H$-matrix of proposed (8, 3) SEC-DED and SEC-DED-DAEC codes}
\label{fig3}
\end{figure}
The $H$-matrix of (8, 3) SEC-DED-DAEC code is provided in Fig. \ref{fig3}. This matrix contains $d_1$ to $d_3$ data columns and $c_1$ to $c_5$ parity columns. 
\begin{figure}[]
\centering
\[
H=
  \left[ {\begin{array}{ccccccccc}
0	1	1	0	1	0	0	0	0\\
1	0	1	0	0	1	0	0	0\\
0	1	0	1	0	0	1	0	0\\
1	1	0	1	0	0	0	1	0\\
1	0	1	1	0	0	0	0	1\\
  \end{array} } \right]
\]
\caption{$H$-matrix of proposed (9, 4) SEC-DED and SEC-DED-DAEC codes}
\label{fig4}
\end{figure}
The $H$-matrix of (9, 4) SEC-DED-DAEC code is provided in Fig. \ref{fig4}. In this matrix consists of 4 data columns and 5 parity columns.
\begin{figure}[]
\centering
\[
H=
  \left[ {\begin{array}{ccccccccccc}
0	0	1	1	0	1	0	0	0	0	0\\
1	1	0	1	0	0	1	0	0	0	0\\
0	1	0	1	1	0	0	1	0	0	0\\
1	0	1	0	1	0	0	0	1	0	0\\
1	0	0	0	0	0	0	0	0	1	0\\
0	1	1	0	1	0	0	0	0	0	1\\
  \end{array} } \right]
\]
\caption{$H$-matrix of proposed (11, 5) SEC-DED and SEC-DED-DAEC codes}
\label{fig5}
\end{figure}
The $H$-matrix of (11, 5) SEC-DED-DAEC code is provided in Fig. \ref{fig5} which consists of 5 data columns and 6 parity columns.
\begin{figure}[]
\centering
\[
H=
  \left[ {\begin{array}{ccccccccccccc}
0	1	0	0	1	1	0	1	0	0	0	0	0\\
0	0	1	1	0	1	0	0	1	0	0	0	0\\
1	1	0	1	0	1	1	0	0	1	0	0	0\\
0	1	1	0	1	0	1	0	0	0	1	0	0\\
1	0	1	0	0	0	0	0	0	0	0	1	0\\
1	0	0	1	1	0	1	0	0	0	0	0	1\\
  \end{array} } \right]
\]
\caption{$H$-matrix of proposed (13, 7) SEC-DED and SEC-DED-DAEC codes}
\label{fig6}
\end{figure}
The $H$-matrix of (13, 7) SEC-DED-DAEC code is provided in Fig. \ref{fig6} where $d_1$-$d_7$ are data columns and $c_1$-$c_6$ are parity columns.
\begin{figure}[]
\centering
\[
H=
  \left[ {\begin{array}{cccccccccccccccccccccccc}
1	0	1	0	0	0	1	0	1	0	1	0	0	1	1	0	1	0	0	0	0	0	0	0\\
0	1	0	0	0	0	0	1	0	0	0	1	1	0	1	0	0	1	0	0	0	0	0	0\\
0	0	0	0	1	1	1	0	0	0	1	0	1	0	1	1	0	0	1	0	0	0	0	0\\
0	0	0	0	0	0	0	0	0	1	1	1	0	1	0	1	0	0	0	1	0	0	0	0\\
1	0	0	1	1	0	1	1	0	0	0	0	0	0	0	0	0	0	0	0	1	0	0	0\\
0	1	1	1	0	1	0	0	0	1	0	1	0	0	0	0	0	0	0	0	0	1	0	0\\
0	0	1	0	1	0	0	1	1	1	0	0	0	0	0	0	0	0	0	0	0	0	1	0\\
1	1	0	1	0	1	0	0	1	0	0	0	1	1	0	1	0	0	0	0	0	0	0	1\\
\end{array} } \right]
\]
\caption{$H$-matrix of proposed (24, 16) SEC-DED and SEC-DED-DAEC codes}
\label{fig7}
\end{figure}
The $H$-matrix of (24, 16) SEC-DED-DAEC code is provided in Fig. \ref{fig7}. It has 16 data columns and 8 numbers of parity columns.
\subsection{Encoding and Decoding Techniques}
In this section, encoding an decoding processes of proposed $(8, 3)$ SEC-DED and SEC-DED-DAEC code has been described.\\

\textbf{{2.2.1 Encoding Process}}\\

The parity bits are collaborated with the data bits and form the codeword in the encoding process. The equations to generate check-bits of proposed $(8, 3)$ SEC-DED and SEC-DED-DAEC code are in the following. \\
\begin{eqnarray}
\label{equ2}
c_1 & =  & d_1 \oplus d_2  \nonumber\\
c_2 & =  & d_2  \nonumber\\
c_3 & =  & d_2 \oplus d_3 \nonumber\\
c_4 & = & d_1 \oplus d_3\nonumber\\
c_5 & = & d_1 \oplus d_3 \nonumber\\
\end{eqnarray}\\

\textbf{{2.2.2 Decoding Process}}\\

Decoding technique has two parts-a) syndrome computation and b) error correction logic. In the first part, error detection can be done by calculating the syndrome value. No error in received codeword is indicated if the syndrome value is zero $(SY = 0)$ else there are some bit-errors. There are some typo errors in Error correction logic subsection in ref. \cite{Tripathi2019}. It is rectified here and the modified error correction logic has been described in the following.\\ 
The error can be corrected by using the error correction block. For single error in one of the data bits the syndrome corresponds to one of the data column. In case of double adjacent errors in $n^{th}$ and $(n+1)^{th}$ bits the corresponding syndrome is modulo-2 sum of nth and $(n+1)^{th}$ columns of $H$-matrix. Finally, the error pattern block compares the double adjacent error syndromes and single error syndrome using 2-input OR (OR2) gates to confirm the occurrence of error in $n^{th}$ bit. If error occurs in the $n^{th}$ bit, then output of OR2 gate is 1 and error correction is done by 2-input XOR (XOR2) logic, which takes $n^{th}$ bit and output of OR2 gate as inputs to produce corrected version of data stored in $n^{th}$ position of codeword. 
\subsection{Calculation of parity-bits}
The main aim of the proposed codes is to minimize the number of 1's in each row and column of the $H$-matrix. The improvement in delay is occurred by minimizing the number of ones in the row of the matrix. The equation (\ref{equ6}) in ref. \cite{Tripathi2019} has been corrected in this section. For weight, $w$=3, the minimum number of parity bits $(n-k)$ are calculated by considering approximate value from the equation (\ref{equ6}). 
\begin{equation}
\label{equ6}
(n-k)\geq(\sqrt{1+2.5k}+1.90)
\end{equation}
The equation (\ref{equ6}) is applicable for SEC-DED-DAEC codes but there is a limitation. This proposed equation is suitable up to 8-bit SEC-DED-DAEC codes. The number of parity bits required for a specific number of data bits are presented in Table \ref{tabmin}. 
\begin{table}[]
\caption{Parity bits required}
\label{tabmin} 
\centering
\resizebox{!}{0.1\textheight}{%
\begin{tabular}{|c|c|c|}
\hline
Codec & Data bit ($k$) & \begin{tabular}[c]{@{}c@{}} Number of \\ parity bit ($P$)\end{tabular} \\ \hline
(8, 3)   & 3            & 5                                                                        \\ \hline
(9, 4)   & 4            & 5                                                                        \\ \hline
(11, 5)  & 5            & 6                                                                        \\ \hline
(12, 6)  & 6            & 6                                                                        \\ \hline
(13, 7)  & 7            & 6                                                                        \\ \hline
(14, 8)  & 8            & 6                                                                        \\ \hline
\end{tabular}
}
\end{table}
\section{Logic gate estimation of complexity analysis}
This section presents the logic gate estimation of complexity analysis which consists of area complexity and critical path delay.
\subsection{Area complexity}
Area complexity in terms of logic gates of proposed and existing SEC-DED and SEC-DED-DAEC codes are presented in Table \ref{tabarea}. Proposed codes require lesser number logic gates compared to other existing codes. Also the area complexity comparison of existing and proposed codes has been presented in terms of 2-input NAND (NAND2) gates.
\begin{table}[]
\caption{Area complexity comparison of proposed codes and existing codes}
\label{tabarea} 
\centering
\begin{tabular}{|c|l|c|c|c|c|c|}
\hline
Codec & \multicolumn{1}{c|}{Schemes} & XOR2 & AND2 & OR2 & NOT & Equivalent NAND2 \\ \hline
\multirow{7}{*}{\begin{tabular}[c]{@{}c@{}}Existing \\ SEC-DED\end{tabular}} & Alabady (9, 4) \cite{Alabady20182}, \cite{Alabady20183} & 31 & 16 & - & 4 & 160 \\ \cline{2-7} 
 & Alabady (9, 4) \cite{Alabady20181} & 15 & 16 & - & 12 & 104 \\ \cline{2-7} 
 & Adalid (8, 4) \cite{Adali2016}, \cite{Adalid2016} & 27 & 17 & 3 & 5 & 156 \\ \cline{2-7} 
 & Hsiao (13, 8) \cite{Hsiao1970} & 51 & 32 & - & 16 & 284 \\ \cline{2-7} 
 & Hamming (13, 8) \cite{Hamming1950} & 59 & 32 & - & 14 & 314 \\ \cline{2-7} 
 & Cha, Yoon (13, 8) \cite{Cha2012} & 58 & 32 & - & 13 & 309 \\ \cline{2-7} 
 & Adalid (16, 8) \cite{Adali2016}, \cite{Adalid2016} & 55 & 25 & 7 & 1 & 292 \\ \hline
\multirow{6}{*}{\begin{tabular}[c]{@{}c@{}}Proposed \\ SEC-DED\end{tabular}} & Proposed (8, 3) & 16 & 6 & - & - & 76 \\ \cline{2-7} 
 & Proposed (9, 4) & 23 & 8 & - & - & 108 \\ \cline{2-7} 
 & Proposed (11, 5) & 29 & 10 & - & - & 136 \\ \cline{2-7} 
 & Proposed (13, 7) & 43 & 14 & - & - & 200 \\ \cline{2-7} 
 & Proposed (14, 8) & 50 & 16 & - & - & 232 \\ \cline{2-7} 
 & Proposed (24, 16) & 120 & 32 & - & - & 544 \\ \hline
\multirow{2}{*}{\begin{tabular}[c]{@{}c@{}}Existing\\ SEC-DED-DAEC\end{tabular}} & Ming (22, 16) \cite{Ming2011} & 112 & 235 & 31 & 120 & 1131 \\ \cline{2-7} 
 & Dutta (22, 16) \cite{Dutta2007} & 106 & 235 & 31 & 126 & 1113 \\ \hline
\multirow{6}{*}{\begin{tabular}[c]{@{}c@{}}Proposed \\ SEC-DED-\\ DAEC\end{tabular}} & Proposed (8, 3) & 20 & 24 & 5 & - & 143 \\ \cline{2-7} 
 & Proposed (9, 4) & 27 & 33 & 7 & - & 195 \\ \cline{2-7} 
 & Proposed (11, 5) & 34 & 42 & 9 & - & 247 \\ \cline{2-7} 
 & Proposed (13, 7) & 48 & 60 & 13 & - & 351 \\ \cline{2-7} 
 & Proposed (14, 8) & 55 & 69 & 15 & - & 403 \\ \cline{2-7} 
 & Proposed (24, 16) & 127 & 141 & 31 & - & 883 \\ \hline
\end{tabular}
\end{table}
\subsection{Critical path delay}
Critical path delay of proposed and existing SEC-DED and SEC-DED-DAEC codes are provided in Table \ref{tabcric}. It has been observed that the performance of proposed codes is better than related existing SEC-DED and SEC-DED-DAEC codes.   
\begin{table}[]
\caption{Critical path delay comparison of proposed codes and existing codes}
\label{tabcric} 
\centering
\begin{tabular}{|c|l|c|c|c|c|c|}
\hline
Codec & \multicolumn{1}{c|}{Schemes} & XOR2 & AND2 & OR2 & NOT & \begin{tabular}[c]{@{}c@{}}Equivalent \\ NAND2\end{tabular} \\ \hline
\multirow{7}{*}{\begin{tabular}[c]{@{}c@{}}Existing \\ SEC-DED\end{tabular}} & Alabady (9, 4) \cite{Alabady20182}, \cite{Alabady20183} & 8 & 4 & - & 1 & 41 \\ \cline{2-7} 
 & Alabady (9, 4) \cite{Alabady20181} & 4 & 4 & - & 2 & 26 \\ \cline{2-7} 
 & Adalid (8, 4) \cite{Adali2016}, \cite{Adalid2016} & 9 & 4 & - & 2 & 46 \\ \cline{2-7} 
 & Hsiao (13, 8) \cite{Hsiao1970} & 10 & 4 & - & 1 & 49 \\ \cline{2-7} 
 & Hamming (13, 8) \cite{Hamming1950} & 20 & 4 & - & 1 & 89 \\ \cline{2-7} 
 & Cha, Yoon (13, 8) \cite{Cha2012} & 18 & 4 & - & 1 & 81 \\ \cline{2-7} 
 & Adalid (16, 8) \cite{Adali2016}, \cite{Adalid2016} & 10 & 3 & 7 & 1 & 68 \\ \hline
\multirow{6}{*}{\begin{tabular}[c]{@{}c@{}}Proposed \\ SEC-DED\end{tabular}} & Proposed (8, 3) & 4 & 2 & - & - & 20 \\ \cline{2-7} 
 & Proposed (9, 4) & 6 & 2 & - & - & 28 \\ \cline{2-7} 
 & Proposed (11, 5) & 6 & 2 & - & - & 28 \\ \cline{2-7} 
 & Proposed (13, 7) & 10 & 2 & - & - & 44 \\ \cline{2-7} 
 & Proposed (14, 8) & 10 & 2 & - & - & 44 \\ \cline{2-7} 
 & Proposed (24, 16) & 16 & 2 & - & - & 68 \\ \hline
\multirow{2}{*}{\begin{tabular}[c]{@{}c@{}}Existing SEC-DED-\\ DAEC\end{tabular}} & Ming (22, 16) \cite{Ming2011} & 22 & 5 & 2 & 1 & 105 \\ \cline{2-7} 
 & Dutta (22, 16) \cite{Dutta2007} & 18 & 5 & 2 & 1 & 89 \\ \hline
\multirow{6}{*}{\begin{tabular}[c]{@{}c@{}}Proposed SEC-DED-\\ DAEC\end{tabular}} & Proposed (8, 3) & 8 & 3 & 2 & - & 44 \\ \cline{2-7} 
 & Proposed (9, 4) & 10 & 3 & 2 & - & 52 \\ \cline{2-7} 
 & Proposed (11, 5) & 11 & 3 & 2 & - & 56 \\ \cline{2-7} 
 & Proposed (13, 7) & 15 & 3 & 2 & - & 72 \\ \cline{2-7} 
 & Proposed (14, 8) & 15 & 3 & 2 & - & 72 \\ \cline{2-7} 
 & Proposed (24, 16) & 23 & 3 & 2 & - & 104 \\ \hline
\end{tabular}
\end{table}
The critical path delays of proposed codes are compared to the Alabady et al. code \cite{Alabady20181}, Adalid et al. code \cite{Adali2016}, Hsiao code \cite{Hsiao1970}, Hamming code \cite{Hamming1950}, Cha-Yoon code \cite{Cha2012} and Ming et al. code \cite{Ming2011}.
\section{Synthesis results}
The proposed SEC-DED and SEC-DED-DAEC codes have been represented in Verilog hardware description language (HDL). All codes have been simulated and synthesized in ASIC platform using Cadence based Genus synthesis solution (TSMC18) tool. The The ASIC-based synthesis results in terms of area, power, delay, power delay product (PDP), power area product (PAP) and cost (product of area, power and delay) of proposed and existing SEC-DED and SEC-DED-DAEC codes have been presented in Table \ref{tabasic}. \\
\begin{table}[]
\caption{ASIC synthesis results of proposed codes and existing codes}
\label{tabasic} 
\centering
\begin{tabular}{|c|l|c|c|c|c|c|c|}
\hline
Codec & \multicolumn{1}{c|}{Schemes} & Area & Power & Delay & PDP & PAP & Cost \\ \hline
\multirow{7}{*}{\begin{tabular}[c]{@{}c@{}}Existing\\ SEC-DED\end{tabular}} & Alabady (9, 4) \cite{Alabady20182}, \cite{Alabady20183} & 592.11 & 59.36 & 365.3 & 21.69 & 0.04 & 0.013 \\ \cline{2-8} 
 & Alabady (9, 4) \cite{Alabady20181} & 475.66 & 38.84 & 282.4 & 10.97 & 0.02 & 0.005 \\ \cline{2-8} 
 & Adalid (8, 4) \cite{Adali2016}, \cite{Adalid2016} & 708.54 & 70.67 & 365.3 & 25.82 & 0.05 & 0.018 \\ \cline{2-8} 
 & Hsiao (13, 8) \cite{Hsiao1970}& 1293.95 & 155.94 & 302.9 & 47.24 & 0.20 & 0.061 \\ \cline{2-8} 
 & Hamming (13, 8) \cite{Hamming1950}& 1204.17 & 145.41 & 328.9 & 47.83 & 0.18 & 0.058 \\ \cline{2-8} 
 & Cha, Yoon (13, 8) \cite{Cha2012} & 1200.84 & 159.58 & 365.3 & 55.55 & 0.19 & 0.070 \\ \cline{2-8} 
 & Adalid (16, 8) \cite{Adali2016}, \cite{Adalid2016} & 1380.45 & 180.83 & 423.5 & 76.58 & 0.25 & 0.106 \\ \hline
\multirow{6}{*}{\begin{tabular}[c]{@{}c@{}}Proposed \\ SEC-DED\end{tabular}} & Proposed (8, 3) & 382.55 & 34.16 & 362.4 & 12.37 & 0.01 & 0.005 \\ \cline{2-8} 
 & Proposed (9, 4) & 542.22 & 56.92 & 365.3 & 20.80 & 0.03 & 0.011 \\ \cline{2-8} 
 & Proposed (11, 5) & 708.52 & 69.73 & 341.9 & 23.84 & 0.05 & 0.017 \\ \cline{2-8} 
 & Proposed (13, 7) & 997.93 & 110.38 & 337.5 & 37.25 & 0.11 & 0.037 \\ \cline{2-8} 
 & Proposed (14, 8) & 1150.95 & 138.02 & 292.7 & 40.40 & 0.16 & 0.046 \\ \cline{2-8} 
 & Proposed (24, 16) & 2318.51 & 344.57 & 308.4 & 106.27 & 0.80 & 0.246 \\ \hline
\multirow{2}{*}{\begin{tabular}[c]{@{}c@{}}Existing \\ SEC-DED-DAEC\end{tabular}} & Ming (22, 16) \cite{Ming2011} & 2877.34 & 486.99 & 467.4 & 227.62 & 1.40 & 0.655 \\ \cline{2-8} 
 & Dutta (22, 16) \cite{Dutta2007} & 2920.58 & 461.16 & 429.3 & 197.98 & 1.35 & 0.578 \\ \hline
\multirow{6}{*}{\begin{tabular}[c]{@{}c@{}}Proposed \\ SEC-DED\\ -DAEC\end{tabular}} & Proposed (8, 3) & 555.52 & 61.18 & 273.6 & 16.74 & 0.03 & 0.009 \\ \cline{2-8} 
 & Proposed (9, 4) & 751.77 & 87.33 & 289.7 & 25.30 & 0.07 & 0.019 \\ \cline{2-8} 
 & Proposed (11, 5) & 908.12 & 115.75 & 502.8 & 58.20 & 0.11 & 0.053 \\ \cline{2-8} 
 & Proposed (13, 7) & 1343.87 & 177.40 & 482.7 & 85.63 & 0.24 & 0.115 \\ \cline{2-8} 
 & Proposed (14, 8) & 1543.44 & 238.85 & 433.8 & 103.61 & 0.37 & 0.160 \\ \cline{2-8} 
 & Proposed (24, 16) & 2826.59 & 619.47 & 434.4 & 269.10 & 1.75 & 0.761 \\ \hline
\end{tabular}
\end{table}
\section{Conclusion}
In this paper, SEC-DED and SEC-DED-DAEC codes have been proposed for different message lengths. These SEC-DED and SEC-DED-DAEC codes have been designed and implemented based on ASIC platform. Performance of our design has been analyzed in terms of area, power, delay, PDP, PAP and cost. The estimation of logic gates for proposed and existing SEC-DED and SEC-DED-DAEC codes are provided. Our proposed design is faster and power efficient than other related designs. 
\section*{Acknowledgment}
Authors are thankful for research grant received from RUSA 2.0 of Jadavpur University. Authors would also like to thank to SMDP-C2SD, Jadavpur University for Cadence simulation software.

\end{document}